# Intrinsically Correct Sorting in Cubical Agda

Cass Alexandru*
c.alexandru@cs.rptu.de
RPTU Kaiserslautern-Landau
Kaiserslautern, Germany

Vikraman Choudhury
vikraman.choudhury@unibo.it
Università di Bologna & Inria OLAS
Bologna, Italy

Jurriaan Rot
j.rot@cs.ru.nl
Radboud University Nijmegen
Nijmegen, The Netherlands

Niels van der Weide
niels.vanderweide@ru.nl
Radboud University Nijmegen
Nijmegen, The Netherlands

## Abstract

The paper "Sorting with Bialgebras and Distributive Laws" by Hinze et al. uses the framework of bialgebraic semantics to define sorting algorithms. From distributive laws between functors they construct pairs of sorting algorithms using both folds and unfolds. Pairs of sorting algorithms arising this way include insertion/selection sort and quick/tree sort.

We extend this work to define intrinsically correct variants in cubical Agda. Our key idea is to index our data types by multisets, which concisely captures that a sorting algorithm terminates with an ordered permutation of its input list. By lifting bialgebraic semantics to the indexed setting, we obtain the correctness of sorting algorithms purely from the distributive law.

## 1 Introduction

Sorting is one of the most widely studied algorithmic problems in computer science. As a consequence, not only a wide variety of sorting algorithms have been developed, but there also have been numerous efforts in verifying these algorithms. These verification efforts range from early non-trivial examples of verified functional algorithms (e.g., [Appel 2017; Nipkow et al. 2021]) to full verification of industrial code [de Gouw et al. 2019].

***Intrinsic and extrinsic verification.*** The most common way of verifying sorting and other algorithms is by what we refer to as *extrinsic* verification. In this approach, the programmer starts by implementing the algorithm in their language of choice. Afterwards, they reason about the algorithm and prove its correctness. The (partial) correctness specification is thus a separate entity from the algorithm. In the case of sorting, one implements a sorting algorithm, and designs a predicate that a list is ordered, a predicate that one list is a permutation of another, and then proves for the algorithm at hand that the output list is ordered and a permutation of the input.

An alternative approach is to directly encode the properties of orderedness and element-preservation in the definitions of the type of the program and in the types of its underlying data structures. Here the programmer would use more expressive data structures, such as ordered lists, to express the correctness of their program in the type. This approach is referred to as *intrinsic verification*. Intrinsic verification may avoid code duplication that occurs in extrinsic verification, because the predicates one defines mirror the recursive structure of the data they predicate over, and the proofs mirror the recursive structure of the programs. However, encoding the specification in the type can be a challenge for intrinsic verification. For sorting algorithms, the first part is to encode orderedness of output lists, which is addressed in [McBride 2014].

We encode both orderedness *and* element-preservation intrinsically, expressing the specification of sorting entirely in the type of the programs. To express element-preservation, our key idea is to used indexed types. More specifically, types are indexed by multisets, capturing intuitively the elements that occur in lists. By using finite multisets, this approach not only allows us to naturally encode orderedness and element-preservation, but also to prove termination. Our development is in cubical Agda, and we represent finite multisets as a quotient inductive type [Univalent Foundations Program 2013] following [Choudhury and Fiore 2023].

***Sorting via bialgebras.*** To further minimize structural proof-program duplication, we define our sorting algorithms in a sophisticated way that allows us to obtain two intrinsically verified algorithms for the price of one shared non-recursive "business logic". More specifically, we construct intrinsically verified versions of sorting algorithms that are defined using the categorical framework of *bialgebraic semantics*, as proposed in [Hinze et al.

---

*first author in terms of contributions

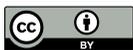

2012]. In this framework, based on the seminal work of Turi and Plotkin on the use of bialgebras in operational semantics [Turi and Plotkin 1997], it is shown that various sorting algorithms arise from *distributive laws* between functors. Distributive laws are extended to full sorting algorithms in two ways, using both folds (induction) and unfolds (coinduction).

For instance, both insertion sort and selection sort arise in this way from the "swap" distributive law that takes a pair of elements and puts them in the right order. In the current paper, we show that intrinsic verification of these algorithms reduces to the verification of this distributive law, thereby focusing only on the key business logic behind the two algorithms. We then show how to extend this to the more complex examples of quicksort, treesort and heapsort, which also arise from distributive laws (which are, accordingly, a bit more complex). Altogether, we thus show how to define intrinsic correctness of sorting algorithms in cubical Agda, and prove correctness of various sorting algorithms as generated from a distributive law in the framework of bialgebraic semantics.

**Contributions.** We summarise the main contributions of this paper as follows.
- We formulate intrinsic correctness of sorting algorithms in cubical Agda, by indexing datatypes of lists by a quotient inductive type of finite multisets.
- We show that the finite multiset index acts as a termination measure which enables us to define maps into an inductive datatype via iteration.
- We define intrinsically correct versions of insertion-, selection-, heap-, tree- and quicksort by verifying only their non-recursive business logic in the form of distributive laws, à la [Hinze et al. 2012].
- We revisit finite multiset indexing at the more semantic level of Set-based coalgebras, showing how ordered (finite) lists arise as a final coalgebra in this setting.

This paper has an accompanying Agda formalization [Alexandru et al. 2024], of which we include excerpts in the text.

## 2 Overview

In this section we describe the main problem addressed in this paper: how to formulate and prove intrinsic correctness of sorting algorithms generated from a distributive law. We start by recalling the analysis by Hinze et al. [2012], set in Haskell, of how a distributive law forms the shared business logic of insertion and selection sort (Sections 2.1 to 2.3). We then move on to the definitions and proofs necessary to define an intrinsically correct version of this algorithm (Section 2.4), thereby sketching our overall approach. Throughout this section, we fix a type $A$ of elements to be sorted, with a total ordering $\leq$ on it.

We analyse the recursion behaviour of insertion and selection sort through the lens of algebra and coalgebra [Jacobs 2016]. To do this, we briefly recall how these category-theoretical notions apply to structured (co)recursion.

Recursive datatypes have a shape given by a non-recursive *base functor* [Bird and de Moor 1997]. For example, the type of lists, *List*, is the recursive datatype for the base functor $L$:

**data** $List = [] \mid A : List$
**data** $L \, r \;\; = \star \mid (A, r)$

In our introductory setting of the lazy functional programming language Haskell, recursive datatypes are the carriers of both the initial algebra and final coalgebra for this functor [Smyth and Plotkin 1982]. We call them inductive when referring to them as carriers of the initial algebra, and coinductive when referring to them as carriers of the final coalgebra. To make this distinction explicit, we introduce the type synonym $\mu L$ for *List* as an inductive datatype, and $\nu O$ for *List* as coinductive. We also introduce a type synonym for the base functor $L$ when we use it in the type of coalgebras.

**type** $\mu L \;\; = List$
**type** $\nu O \;\; = List$
**type** $O \, r = L \, r$

One can define maps *out of* inductive datatypes as *folds* of algebras. Intuitively, folds are bottom-up traversals of a term that replace its constructors with functions that return values of some type $x$. The business logic for replacing the constructors of a datatype with base functor $F$ is provided as a map of type $F\,x \to x$. Such a map is called an $F$-algebra with carrier $x$. The inductive datatype $\mu F$ is the carrier of the initial $F$-algebra $in :: F\,\mu F \to \mu F$. This algebra replaces the constructors of $\mu F$ with themselves. The type of the fold for $L$ is:

$fold :: (L\,x \to x) \to (\mu L \to x)$

One can define maps *into* into coinductive datatypes as *unfolds* of coalgebras. Intuitively, coalgebras are functions that, given a seed value, produce one layer of a coinductive datatype, with new seeds at the recursive positions. Successively applying such a coalgebra to the new seeds until only leaves of the datatype are produced yields possibly infinite trees of the shape given by the base functor. An $F$-coalgebra with carrier $x$ is a map of type $x \to F\,x$. The coinductive datatype $\nu F$ is the carrier of the final $F$-coalgebra $out :: \nu F \to F\,\nu F$. This coalgebra exposes one layer of recursion of the codatatype. The type of the unfold for $O$ is:

$unfold :: (x \to O\,x) \to (x \to \nu O)$

The type of a sorting algorithm is a map *from* a list *to* a list. To emphasise that we treat the input inductively and the output coinductively, we write its type as

$$sort :: \mu L \to \nu O \,.$$

Note that maps of this type can be constructed *out of* the domain using *fold*, or *into* the codomain using *unfold*. Both methods are used in the bialgebraic framework, yielding two sorting algorithms from a single piece of business logic (a distributive law). We now proceed to analyse the recursion behavior of insertion and selection sort using folds and unfolds.

### 2.1 Insertion sort

Insertion sort is a list traversal that successively inserts elements at the correct place into the output list, starting with an empty list. As such we can express it as a fold of the algebra *insert*:

*insert* :: $L\ \nu O \to \nu O$
*insert* $\star = [\,]$
*insert* $(a, [\,]) = a : [\,]$
*insert* $(a, b : r')$
  | $a \leq b$ = $a : b : r'$
  | otherwise = $b :$ *insert* $(a, r')$

*insertSort* = *fold insert*

In turn, the algebra *insert* can be defined recursively using an *O*-coalgebra, as a variant of an *unfold*. This coalgebra has as seed type a pair of an element to be inserted and a list it is to be inserted into. At each step, it outputs an element and continues with a new seed, returning early when the element to be inserted is a lower bound to the rest of the list.

This type of recursion is a modified version of an unfold, known as an *apomorphism* [Vene and Uustalu 1998]. The ability to return early is captured by the addition of $\nu O + \cdot$ in the carrier of the coalgebra.

*apo* :: $(\ x\ \to O\ (\nu O +\ x\ )) \to x \to \nu O$
*ins* :: $(L\ \nu O \to O\ (\nu O + L\ \nu O))$
*ins* $\star = \star$
*ins* $(a, [\,])$   = $(a, $ *Left* $[\,])$
*ins* $(a, b : r')$
  | $a \leq b$ = $(a, $ *Left* $(b : r'))$
  | otherwise = $(b, $ *Right* $(a, r'))$
*insert* = *apo ins*

We have now expressed the recursion behaviour of insertion sort in terms of folds and unfolds. The outer layer is expressible as a fold, the inner as early-return-enabled unfold (an apomorphism), supplied with a non-recursive coalgebra:

*insertSort* = *fold* (*apo ins*)

### 2.2 Selection sort

We continue with the analysis of the recursion behaviour of selection sort to reduce its business logic to a non-recursive step, similar to the case of insertion sort. We then lay out how those two non-recursive business logics can be synthesised from a single *shared*, and thus even more highly distilled, non-recursive business logic.

Selection sort can be defined using an *O*-coalgebra. This coalgebra has a list as seed. At each step, it produces an element of the output, and continues with the list this element was removed from as the new seed. As such, we can express it as an unfold.

*select* :: $\mu L \to O\ \mu L$
*select* $[\,] = \star$
*select* $(a : r) = $ **case** *select r* **of**
  $\star$      $\to (a, r)$
  $(b, r') \to$ **if** $a \leq b$
    **then** $(a, r)$
    **else** $(b, a : r')$
*selectSort* = *unfold select*

Dually to the case of insertion sort, the coalgebra *select* can itself be defined recursively, using an *L*-algebra. This algebra traverses the input to produce the least element and the list it was extracted from. To return this list, it needs to have access to the tail of the list as well as the result of its recursive application to it. This type of recursion is a modified version of a fold, known as a *paramorphism* [Meertens 1992]. The access to the original subterm(s) is encoded by an additional $\mu L$ argument in the carrier of the algebra.

*para* :: $(L\ (\mu L \times\ x\ ) \to\ x\ ) \to \mu L \to x$
*sel*  :: $(L\ (\mu L \times O\ \mu L) \to O\ \mu L)$
*sel* $\star = \star$
*sel* $(a, (r, \star)) = (a, r)$
*sel* $(a, (r, (b, r')))$
  | $a \leq b$ = $(a, r)$
  | otherwise = $(b, a : r')$
*select* = *para sel*

We have again entirely refactored two layers of explicit recursion into structured recursion, this time for selection sort. The outer layer of recursion is expressible as an unfold, the inner as a fold with access to the original input (a paramorphism), supplied with a non-recursive algebra:

*selectSort* = *unfold* (*para sel*)

It remains to distill a shared non-recursive business logic from the definitions of *ins* and *sel*, which justifies labeling the two algorithms as dual.

### 2.3 Synthesis

The shared business logic of *ins* and *sel* is the following parametrically polymorphic function *swap*.

*swap* :: $L\ (x \times O\ x) \to O\ (x + L\ x)$
*swap* $(a, (r, \star)) = (a, $ *Left* $r)$
*swap* $(a, (r, (b, r')))$

```
| a ≤ b      = (a, Left r)
| otherwise  = (b, Right (a, r'))
```

We can reconstruct *ins* from it using the tupling operation $\langle f, g \rangle \, x = (f \, x, g \, x)$ applied to *id* and *out* :: $\nu O \to O \, \nu O$. We lift this to the codomain $L \, \nu O$ of *ins* by $L_1$, the action of the functor $L$ on maps, and postcompose it to *swap*.

$ins = swap \circ L_1 \, \langle id, out \rangle$

Dually, we can reconstruct *sel* using the cotupling operation $[\cdot, \cdot] :: (b \to a) \to (c \to a) \to (b + c) \to a$ applied to *id* and *in* :: $L \, \mu L \to \mu L$. We lift this to the domain $O \, \mu L$ of *sel* by $O_1$, and precompose it to *swap*.

$sel = O_1 \, [id, in] \circ swap$

In summary, from one non-recursive business logic, we obtain two algorithms with quite different recursion behaviour. This shared business logic is given by the function $swap :: L \, (x \times O \, x) \to O \, (x + L \, x)$, which is parametrically polymorphic in $x$. We can phrase this as a *distributive law*. Here a distributive law between functors $F$ and $G$ is a parametrically polymorphic function of type $F \, (G \, x) \to G \, (F \, x))$. If one defines $F_+ x = x + F \, x$ and $F_\times x = x \times F \, x$, then *swap* is equivalent to a function of type $L_+(O_\times x) \to O_\times(L_+ x)$. This is proved by Hinze et al. [2012, Appendix A] as a specialization of a result by Lenisa et al. [2000].

The two algorithms one obtains from it, by a fold of an unfold and an unfold of a fold respectively, are called its *bialgebraic semantics*. As we have reconstructed the two algorithms we started with from the distributive law *swap*, we now refer to them as the *bialgebraic* sorting algorithm insertion/selection sort. This concludes our summary of the bialgebraic approach to sorting [Hinze et al. 2012].

### 2.4 Intrinsically correct bialgebraic sorting

Our goal in this paper is to express an intrinsically verified version of bialgebraic sorting algorithms. As such, our setting changes from a partial to a *total* programming language, for which we chose cubical Agda [Vezzosi et al. 2019]. We distinguish Agda code from Haskell code with coloured syntax highlighting.

The first issue to address is that in a total language, inductive and coinductive datatypes no longer coincide. Inductive datatypes are well-founded trees, while coinductive datatypes are not necessarily well-founded. This is because unfolding coalgebras may, in general, produce infinite data. Consider for example the following function.

$c :: A \to (() \to O \, ())$
$c \, a = \lambda() \to (a, ())$

Partially applied to a value of type $A$, it is an $O$-coalgebra with the unit type as its seed type. At each step it outputs the element $a$ and continues with the single inhabitant of the unit type as the new seed. The unfold of this coalgebra is therefore an infinite stream of $a$'s.

This example showcases that whether an unfold produces finite or infinite data depends entirely on the coalgebra unfolded. All the coalgebras defined in the previous section do in fact produce only finite output, but this fact needs to be encoded in the type system.

Leaving this issue aside for the moment, we consider the specification of sorting. There are two rules of sorting.

1. The output list must be *ordered*, i.e. consecutive elements should be related by $\leq$.
2. The output must be a permutation of the input.

The first property has been verified intrinsically before, e.g. by McBride [2014].

It is not immediately clear how to state the second property intrinsically. It relates the output to the input, so it must be a property of the program, and not solely the datatypes. There is a different, but still extrinsic, way of stating it that lends itself to being stated intrinsically. This reformulation is that mapping a list to the multiset of its elements should be invariant under sorting.

We can state this property intrinsically using *indexed datatypes* [Dybjer 1994], namely:

> Sorting is an index-preserving map between lists and ordered lists indexed by the finite multiset of their elements.

We formulate this as a type in Agda. First, we fix a type A : Type $\ell$ (where $\ell$ is a universe level), with a total order on it. Using a suitable type FMSet for finite multisets, we define the following.

1. An indexed type EIList : FMSet A $\to$ Type $\ell$ of lists indexed by the multiset of their elements;
2. An indexed type OEIList of *ordered* element-indexed lists.

Then, a term inhabiting the following type is an intrinsically correct sorting algorithm.

$$\{g : \mathsf{FMSet}\, A\} \to \mathsf{EIList}\, g \to \mathsf{OEIList}\, g$$

Categorically, index-preserving maps are the arrows in the category of families over a fixed indexed type, in this case FMSet A $\to$ Type $\ell$. We explore this categorical semantics further in Section 6.

We use a *shallow embedding* [Boulton et al. 6 24; Gibbons and Wu 2014] of indexed type families. This means in particular that equality of the multiset type FMSet A is determined by its type's identity type in the host language. This is in contrast to using *setoids* [Barthe et al. 2003], where one internalizes equality as user-defined equivalence relations for each type. This is our motivation for working in cubical Agda – it allows us to specify a type's equality using *higher inductive types*. We return to this in Section 3.

To define sorting using bialgebraic semantics in the setting of types indexed by finite multisets, we do the following.

1. Define base functors O, L : (FMSet A → Type ℓ) → FMSet A → Type ℓ, such that:
   a. EIList is the carrier of the initial L-algebra
   b. OEIList is the carrier of the final O-coalgebra (!)
2. Define a version of the distributive law swap, which by the definitions of L and O is intrinsically correct.

We made a note of Item 1b for two reasons. First, it means we encode orderedness at the level of base functors, not just in the inductive datatype OEIList. Second, because OEIList is inductive, in order to unfold into it we prove that all O-coalgebras terminate after producing only finite output. This means there is a coincidence of the carriers of the initial algebra and final coalgebra. Thus, OEIList is an inductive datatype admitting a coinduction principle.

### Plan of the Paper.

- In Section 3 we motivate our choice of and introduce a quotient inductive type for finite multisets.
- In Section 4 we define insertion/selection sort using bialgebraic semantics in the setting of types indexed by finite multisets. In Section 4.1, we define the base functors L and O. In Section 4.3, we define the unfold for O-coalgebras into OEIList, and explain why this is possible. Finally, in Section 4.4, we define the distributive law.
- In Section 5, we apply the same "lifting to the indexed setting" to the two-step, non-trivial, bialgebraic sorting algorithms quick/treesort and heapsort, also defined in [Hinze et al. 2012]. We obtain intrinsically correct insertion and deletemin operations for search trees and heaps as useful by-products.
- In Section 6 we give a set-theoretic categorical semantics of our construction.
- Related work is discussed in Section 7. We discuss working with types indexed by higher inductive types in Section 8, and conclude in Section 9.

## 3 Finite Multisets in Cubical Agda

One of the crucial ingredients in our development is given by the type FMSet A of finite multisets whose elements come from some type A. We would like to define this type in such a way that propositional equality coincides with equality of finite multisets. In addition, we would like to be able to use pattern matching on this type.

We use *quotient inductive types* to acquire the desired type. While inductive types allow us to define data types by specifying their constructors, quotient inductive types are more expressive. We use quotient inductive types to define data types by specifying their constructors and their equality. This allows us to define the desired type of finite multisets: the constructors give us the empty multiset and allow us to add elements to multisets, and we say that by swapping two elements, the multiset stays the same.

While quotient types have been studied in various flavours of type theory [Maietti 1998], quotient inductive types became prominent in homotopy type theory [Univalent Foundations Program 2013]. Consequently, even though quotient types are absent in most proof assistants based on dependent type theory, quotient inductive types are present in cubical Agda. For this reason, we use cubical Agda in this paper.

In this section, we discuss the type of finite multisets in more detail. We start by giving a brief introduction to cubical Agda, and quotient inductive types. After that, we describe the type of finite multisets and we give the relevant operations.

### 3.1 Brief Introduction to Cubical Agda

Cubical Agda is a proof assistant based on *cubical type theory* [Cohen et al. 2017], which is a flavour of *homotopy type theory* (HoTT) [Univalent Foundations Program 2013]. Throughout this paper, we use the language of HoTT and we use several ideas of homotopy type theory. In this section, we briefly recall HoTT.

Homotopy type theory is an extension of Martin-Löf type theory (MLTT) [Martin-Löf 1984] that differs from classical MLTT in how objects are identified. In MLTT, we view the identity type $a = b$ as the type of equality proofs between $a$ and $b$. This type is described formally using an introduction rule and an elimination rule. However, the identity type is not fully specified in the sense that the identity of universes and identity types themselves are not determined by these rules. More specifically, one can consistently assume either uniqueness of identity proofs (all proofs of identity are equal themselves) or the univalence axiom, which states that identity of types corresponds to equivalence. In HoTT, we assume the univalence axiom, and thus the identity type has a stronger specification in that foundation.

As a consequence, we also view types in a different way in HoTT. While usually we interpret types in Martin-Löf type theory as sets, we interpret types in homotopy type theory as topological spaces. Terms are elements of a space, and inhabitants of $a = b$ are seen as *paths* from $a$ to $b$. If we have inhabitants $p, q : a = b$, i.e., two paths, then inhabitants of $p = q$ are seen as *homotopies* between the paths $p$ and $q$. In this interpretation, the identity type becomes proof relevant, because it is a known fact in topology that not all paths are homotopic. This is in contrast to the interpretation of types as sets, where identity is viewed as a set of at most one element.

The proof relevant nature of identity, a core concept in homotopy type theory, and thus also in cubical Agda, is witnessed by discerning types by the level at which the homotopical structure becomes "trivial". One can develop a whole hierarchy of homotopy levels, but in this paper, we are only interested in mere propositions and sets.

A type A is called a (mere) *proposition* if for all terms a and b of type A, we have a path a ≡ b. In a proposition, all inhabitants are equal, and one can also show that all paths of a proposition are equal. The only information we can get about them is whether they are inhabited.

A type A is called a *set* if for all terms a and b of type A, the type a ≡ b is a mere proposition. Concretely, this means that there is at most one way in which two elements can be equal, so identity is proof-irrelevant. This coincides with how equality is treated in set theory, and for that reason, we call such types sets.

### 3.2 Finite Multisets as a Quotient Inductive Type

Another core feature of cubical Agda is given by higher inductive types [Univalent Foundations Program 2013] and, what we use in this paper, *quotient inductive types* (QITs). In this section, we introduce quotient inductive types, and demonstrate how to use them to define finite multisets.

To specify a quotient inductive type, one must describe its constructors and its path constructors. Constructors tell us how to construct inhabitants of the type, and the path constructors tell us how to prove equalities. Each quotient inductive type is required to be a set, which is what distinguishes them from the more general higher inductive types. As an example, we have the type of finite multisets due to [Choudhury and Fiore 2023], which is implemented in the standard library of cubical Agda [The agda/cubical development team 2018].

```
data FMSet (A : Type ℓ) : Type ℓ where
  []     : FMSet A
  _::_   : (x : A) → (xs : FMSet A) → FMSet A
  comm   : ∀ {x y xs} → x :: y :: xs ≡ y :: x :: xs
  trunc  : isSet (FMSet A)
```

We have two point constructors. The first one is [], which gives us the empty multiset. The second one is ::, which allows us to add an element from A to some given multiset. These two constructors are the same as what we would use to define the inductive type of lists.

The difference between the type of lists and of finite multisets comes from the path constructors. Whereas the type of lists is fully specified by its point constructors, the type of finite multisets also has path constructors. More specifically, we have a path constructor comm, which identifies the lists x :: y :: zs and y :: x :: zs. This means that the order of the elements does not matter. Note that we can apply the path constructor comm arbitrarily deep in the list by using congruence. More precisely, we have a function congS.

congS : ∀ {B : Type ℓ} → (f : A → B) (p : x ≡ y) → f x ≡ f y

This allows us to construct the following equality.

1 :: 2 :: 3 :: [] ≡⟨ congS (1 ::_) comm ⟩ 1 :: 3 :: 2 :: []

We can also use composition of paths to swap as many elements as desired. Composition has the following type.

$$\_\cdot\_ : x \equiv y \to y \equiv z \to x \equiv z$$

Finally, we also have a constructor trunc, which expresses that our type is a set. This means that for all xs ys : FMSet A, every two inhabitants p and q of type xs ≡ ys are equal.

In Cubical Agda, an elimination principle is generated for HITs that allows one to map out of them by pattern matching on their point- and path constructors. Using this, Choudhury and Fiore [2023, Definition 2.6] define an alternative eliminator for FMSet A, which witnesses its universal property as the free commutative monoid. Given a commutative monoid (M, ⊗) with M a set, and a map f : A → M, there exists a unique map f♯ : FMSet A → M with the following properties:

f♯-nil   : f♯ [] ≡ e
f♯-cons  : ∀ x xs → f♯ (x :: xs) ≡ f x ⊗ f♯ xs
f♯-++    : ∀ xs ys → f♯ (xs ++ ys) ≡ f♯ xs ⊗ f♯ ys

An example application of this, which we use in Section 4.3, is to define a length map by mapping to the commutative monoid of the naturals with addition, by sending elements of type A to 1. Besides the definition of the map itself, this yields the following lemmata:

length     : FMSet A → ℕ
length-[]  : length [] ≡ 0
length-::  : ∀ {x xs} → length (x :: xs) ≡ 1 + length xs
length-++  : ∀ {x y} → length (x ++ y) ≡ length x + length y

## 4 Multiset-indexed Sorting

In this section we discuss how to lift the bialgebraic sorting algorithm insertion/selection sort to the dependently-typed setting. In Section 4.1 we discuss how to encode element-indexing as well as the orderedness invariant for lists at the level of base functors. We define the latter with an "All" predicate transformer [Hermida and Jacobs 1995] for the finite multiset quotient inductive type (QIT). In Section 4.3 we show how the finite multiset index acts as a termination measure which means that the inductive datatype OEIList is the carrier of the final O-coalgebra.

### 4.1 Base Functors for Element-Indexed (Ordered) Lists

Our goal in this section is to define the base functor for lists indexed by the multiset of their elements. Concretely, we define a dependent type FMSet A → Type ℓ. In cubical Agda [Vezzosi et al. 2019], we can define type families indexed by a HIT, based on theory developed by Cavallo and Harper [2019]. We use this to define the desired base functor.

Intrinsically Correct Sorting in Cubical Agda

```
data L (r : FMSet B → Type l) : FMSet B → Type l where
  []   : L r []
  _::_ : {g : FMSet B} (x : B) → (r g) → L r (x :: g)
```

Similarly, we define the datatype EIList of lists indexed by the finite multiset of their elements.

```
data EIList : FMSet B → Type l where
  [] : EIList []
  _::_ : {g : FMSet B} (x : B) → (EIList g) → EIList (x :: g)
```

The same idea is used in both the base functor and the type EIList. Whenever we add an element to a list, then this element is also added to the multiset containing all the elements.

If we have two multisets els and els' and a path p : els ≡ els' then we are able to transport lists xs of type EIList els to lists of type EIList els'. To do so, we use the function subst, and we can see this in action in the following example.

```
ex1 : EIList (1 :: 2 :: [])
ex1 = 2 :: 1 :: [] € subst EIList comm
```

Here € is flipped function application.

The output datatype, in addition to being indexed by its elements, should also contain these elements in ascending order. We make the observation that due to the transitivity of ≤, requiring each element in a list to be smaller than its successor is equivalent to requiring it be smaller than *all* its successors – and that is something we can encode using the multiset index. To encode this however, we first need to define a predicate transformer stating that some predicate holds for all elements of a multiset.

### 4.2 The All Predicate Transformer

Our goal in this section is to define a predicate transformer All [Hermida and Jacobs 1995] that expresses that some predicate holds for all elements in a finite multiset. A first attempt would be to define this transformer more generally. Instead of defining it as a predicate transformer, we define it as a transformer of type families. More specifically, we would use the following indexed inductive type.

```
data All (P : A → Type ℓ') : FMSet A → Type (ℓ ⊔ ℓ') where
  []A   : All P []
  _::A_ : ∀ {x xs} → P x → All P xs → All P (x :: xs)
```

The constructor []A says that a predicate holds for all members of the empty multiset, and the constructor ::A says that if some predicate holds for both x and all elements of xs, then it also holds for all elements of x :: xs.

However, there are a couple of issues with this definition, and we do not use it in the remainder. One requirement that we have of the predicate transformer All, is that whenever some predicate holds for all elements of x :: xs, then it predicate also holds for x and all elements xs. More specifically, we require there to be a function uncons that transforms inhabitants of All P (x :: xs) into pairs of inhabitants of P x and All P (xs). However, Agda does not allow us to pattern match on inhabitants of type All P (x :: xs), which prevents us from defining the desired function. Intuitively, we can see why such a function would be problematic for arbitrary type families. Since we are looking at finite multisets, there could be multiple occurrences of x in some finite multiset xs. If we have an inhabitant of type All P xs, we might have chosen different terms of type P x for different occurrences of x in xs, which would make the function uncons not well-defined.

Instead we define the predicate transformer All by eliminating from FMSet A to hProp ℓ, the homotopy type of propositions. Homotopy types in cubical Agda are *deeply embedded* [Gibbons and Wu 2014] as dependent pairs of a Type and a proof that it is of a respective homotopy level. So, for example, hProp ℓ = Σ[ X ∈ Type ℓ ] isProp X. This is in contrast to a shallow embedding in the structure of Type itself, as is done e.g. in the Arend theorem prover [JetBrains Research 2021].

Since FMSet A is the free commutative monoid (see Section 3), given a predicate P : A → hProp ℓ we can define a map P♯ : FMSet A → hProp ℓ by providing a commutative monoid structure on hProp ℓ, and showing that hProp ℓ is a set ([Univalent Foundations Program 2013, Theorem 7.1.7]). We choose the monoid structure with the unit type as unit, and conjunction as the multiplication operation. Note that this definition is the same as what one would do for finite sets [Frumin et al. 2018, Definition 5.8], except that we do not have to take idempotence into account. With the universal property of FMSet A, we also construct the following operations on the predicate transformer All. The utility function ⟨_⟩ returns the Type ℓ component of the pair hProp ℓ.

```
uncons-A : ∀ {x xs} → All P (x :: xs) → ⟨ P x ⟩ × All P xs
_::-A_   : ∀ {x xs} → ⟨ P x ⟩ → All P xs → All P (x :: xs)
[]-A     : All P []
_++-A_   : ∀ {x y} → All P x → All P y → All P (x ++ y)
mapAll   : {Q : A → hProp ℓ}
           (P⇒Q : ∀ {x}  → ⟨ P x ⟩  → ⟨ Q x ⟩) →
                 ∀ {xs} → All P xs → All Q xs
```

To use the partially applied relation ≤ : A → A → Type ℓ as a *predicate*, in the sense of a map of type A → hProp ℓ, we need an additional premise ∀ (a b) → isProp (a ≤ b) to construct the binary predicate ≤h : A → A → hProp ℓ.

We can now define the datatype of ordered, element-indexed lists OEIList, given by base functor O, by requiring as an additional argument to the _≤::_ constructor a proof that the element to be prepended is smaller than all the elements in the multiset index of its recursive position.

```
data O (r : FMSet A → Type ℓ) : FMSet A → Type ℓ where
  []   : O r []
```

```
_≤::_ : {g : FMSet A} (x : A) → (rg : r g) →
  All (x ≤h_) g → O r (x :: g)

data OEIList : FMSet A → Type ℓ where
  [] : OEIList []
  _≤::_ : {g : FMSet A} (x : A) → (rg : OEIList g) →
    All (x ≤h_) g → OEIList (x :: g)
```

We define the following utility functions which are useful when prepending an element to an ordered list of which we know the head, because if it is smaller than the head, by transitivity it is smaller than all the rest:

```
≤-to-# : a ≤ b → All (b ≤h_) xs → All (a ≤h_) xs
≤-to-# a≤b = mapAll (≤-trans a b _ a≤b)

_≤::#_ : a ≤ b → All (b ≤h_) xs → All (a ≤h_) (b :: xs)
a≤b ≤::# b#c = a≤b ::-A ≤-to-# a≤b b#c
```

### 4.3 Final Coalgebra

Next we must prove that the inductive datatype OEIList is the carrier of the final O-coalgebra. This amounts to defining a map unfoldO that iterates any given O-coalgebra until it produces [], returning its output as a list. Because unfoldO's output is intended to be a list and not a coinductive stream, it needs to be the case that all O-coalgebras terminate after producing only a finite number of elements. We will motivate that this is, in fact, the case, by considering a first attempt at a definition of unfoldO:

```
pattern _≤::_^_bc_ x rg g prf = _≤::_ {g = g} x rg prf

unfoldO : {r : FMSet A → Type ℓ} →
  ( ∀ {g_r} → r g_r → O r g_r) →
    ∀ {g : FMSet A} → r g → OEIList g
unfoldO grow {g} seed with grow seed
unfoldO grow .{[]}      seed | [] = []
unfoldO grow .{x :: g'} seed | x ≤:: seed' ^ g' bc prf =
  (x ≤:: unfoldO grow {g'} seed') prf
```

First, some exposition is called for. The pattern keyword allows us to define a *pattern synonym* for the _≤::_ constructor. We append ^ to the argument in the recursive position to annotate it with its multiset index. Second, we make use of Agda's implementation of with-abstraction [McBride and McKinna 2004] to pattern-match on the result of the application of the coalgebra grow to the seed value. This pattern-match lets us refine the argument g with *dot patterns* [Agda Team 2024] that document the only possible value for it in each case, determined by the value of grow seed.

We can now observe the fact that for any O-coalgebra applied to a seed, the multiset index of the new seed, if produced, is strictly smaller than that of the original. Recall that an O-coalgebra c is an index-preserving map of type ∀ g₂ → r g₂ → O r g₂. Every seed value s to which it can be applied, has some index g. If we take a step with c, we either end up in the [] case of O, or we output some element a and a new seed s'. This value, a ≤:: s', must have index a :: g', which, due to index preservation, must be equal to the original seed index g, i.e. g = a :: g'. Thus, g' has "decreased" with respect to g, by precisely the element output.

Therefore, the finite multiset index of a seed tells us exactly which elements will be output before terminating (though not in what order) when we iteratively apply c to it. This allows every coalgebra to be *simulated* by an element of the inductive datatype OEIList.

Given this exposition, we might expect the termination checker to accept this function definition. However, taking arguments of HITs matched by dot patterns into account when checking termination is in general inconsistent [Pitts 2020]. Therefore, we must justify termination by manually, and we do this by using *well founded induction*. We define a family of functions indexed by FMSet A by induction on the *length* of the index.

We use the length function defined in Section 3 to define a binary relation <l on FMSet A as the inverse image of < on ℕ under length. As the inverse image of a well founded relation, it is again well founded. Consequently, we can define unfoldO using well founded induction, with the proof of <l supplied as an argument to the induction hypothesis.

```
<l-:: : ∀ {x xs} → length xs < length (x :: xs)
<l-:: {x} {xs} = 0 , refl

unfoldO : ( ∀ {g_r} → r g_r → O r g_r) →
          ∀ {g} → r g → OEIList g
unfoldO grow {g = g} = WFI<l.induction step g where
  step : ∀ x → (∀ y → y <l x → (r y → OEIList y)) →
    (r x → OEIList x)
  step g IH seed with grow {g_r = g} seed
  step .{[]}      IH    seed | [] = []
  step .{x :: g'} IH    seed | x ≤:: seed' ^ g' bc prf =
    (x ≤:: IH g' <l-:: seed') prf
```

We note that m < n is defined as Σ[ k ∈ ℕ ] k + (suc m) ≡ n, which is why witnesses of proofs of type xs <l ys are pairs.

### 4.4 Putting It All Together

Now we put the ideas in this section together to define sorting algorithms using distributive laws. We assume that we have a type A together with a total order ≤. More specifically, that ≤ is valued in propositions, reflexive, transitive, and it satisfies totality. We express totality as follows.

$$≤?≥ : (a\ b : A) → (a ≤ b) ⊎ (b ≤ a).$$

To define the distributive law, we use the product and coproduct of indexed types. We construct the products and coproducts pointwise.

$\_+\_\_\times\_ : (l : X \to \text{Type } \ell') (r : X \to \text{Type } \ell'') \to$
$\quad (i : X) \to \text{Type } (\ell' \sqcup \ell'')$
$(l + r)\ i = l\ i \uplus r\ i$
$(l \times r)\ i = l\ i \times r\ i$

Using these operations on indexed types, we define the distributive law that we use for insertion- and selection sort.

swap : {r : FMSet A → Type ℓ} → {g : FMSet A} →
  L (r × O r) g → O (r + L r) g
swap [] = []
swap (a :: (x , [])) = (a ≤:: inl x) []-A
swap (a :: (x , (b ≤:: x') b#x')) with a ≤?≥ b
...| inl a≤b = (a ≤:: inl x)       $ a≤b ≤::# b#x'
...| inr b≤a = (b ≤:: inr (a :: x')) $ b≤a ::-A b#x' €
  subst (O (_ + L _)) comm

Here $ is the operator for function application with negative precedence, € is the *flipped* application operator.

We conclude this section by defining insertion- and selection sort via bialgebraic semantics.

insertSort selectSort : {g : FMSet A} → ElList g → OElList g
insertSort {g = g} =
  foldL (apoO (swap ∘ L₁ (⟨ id , outO ⟩ {g})))
selectSort {g = g} =
  unfoldO (paraL (O₁ ([ id , inL ] {g}) ∘ swap))

## 5 Lifting Treesorts

So far we have seen and verified the naive bialgebraic sorting algorithm insertion/selection sort. However, these are not the only bialgebraic sorting algorithms defined by Hinze et al. [2012]. They also define tree/quicksort and variants of heap sort. We continue by intrinsically verifying these algorithms.

The algorithms treesort and heapsort are defined by building up and then tearing down a binary search tree and heap respectively. Quicksort joins their ranks as such a two-step algorithm if one reifies its call tree as a binary search tree, a realization due to Turner [1995]. Both the building up and the tearing down step can be defined using bialgebraic semantics.

### 5.1 Base Functors

In the first phase of these algorithms we build up binary search trees and heaps respectively. To use bialgebraic semantics we have to encode their orderedness invariants at the level of base functors in FMSet A → Type ℓ.

We have only one choice in how to do this using only the multiset index. That is to add the orderedness invariants as additional constructor arguments using the All predicate transformer (see Section 4.2). We only show the search tree base functor here. The base functor H for heaps is identical except that the two proof arguments are All (x ≤h_) g₁ and All (x ≤h_) g₂.

data S (r : FMSet A → Type ℓ) : FMSet A → Type ℓ where
  leaf : S r []
  _|⌈_⌉|_ : ∀ {g₁ g₂} → (lt : r g₁) → (x : A) → (rt : r g₂) →
    All (_≤h x) g₁ → All (x ≤h_) g₂ → S r (x :: g₁ ++ g₂)

The inductive data type corresponding to S is called STree. We also introduce a pattern synonym for |⌈_⌉| which allows us to to use infix notation while explicitly writing the implicit FMSet A arguments. We append ^ to the arguments in the recursive positions to annotate them with their multiset index.

pattern _^_|⌈_⌉|_^_ lt g₁ x rt g₂ p1 p2 =
  _|⌈_⌉|_ {g₁ = g₁} {g₂ = g₂} lt x rt p1 p2

### 5.2 Final Coalgebra

The intermediate data structures of binary search trees and heaps that we define are the carriers of both the initial algebra and the final coalgebra for their base functor. In fact, these two-step algorithms illustrate the significance of this coincidence even more. We use unfold when building up the intermediate data structure and fold when tearing it down.

We thus have to define unfold for S. As in Section 4.3, we define a preliminary version which will not be accepted by the termination checker, but which will guide us in defining it by well founded induction.

unfoldS : {r : FMSet A → Type ℓ} →
  ( ∀ {g_r} → r g_r → S r g_r) →
    ∀ {g : FMSet A} → r g → Stree g
unfoldS grow {gs} seed with grow seed
unfoldS grow .{[]} seed           | leaf = leaf
unfoldS grow .{x₁ :: g₁ ++ g₂} seed |
  (ls ^ g₁ |⌈ x₁ ⌉| rs ^ g₂) prf₁ prf₂ with
    unfoldS grow {g₁} ls | unfoldS grow {g₂} rs
... | left | right = (left |⌈ x₁ ⌉| right) prf₁ prf₂

In this definition, the arguments in the recursive calls, g₁ and g₂, are not even *structurally* smaller than the original argument, x₁ :: g₁ ++ g₂, so it would be rejected by the termination checker regardless. We therefore need to provide proofs that g₁ <l x₁ :: g₁ ++ g₂ and g₂ <l x₁ :: g₁ ++ g₂, respectively. These are provided by the lemmas sub_l and sub_r, for whose implementation we refer to the formalization [Alexandru et al. 2024]. The reason they, perhaps counterintuitively, respectively take g₂ and g₁ as arguments (and not the other way around), is because these are the witnesses for the *difference* in length between the original argument and that of the recursive calls.

unfoldS : ( ∀ {g_r} → r g_r → S r g_r) →
  ∀ {g} → r g → Stree g

```
unfoldS grow {g = g} = WFI<l.induction step g where
  step : ∀ x → (∀ y → y <l x → (r y → Stree y)) →
    (r x → Stree x)
  step gs IH seed with grow {g_r = gs} seed
  ...| leaf                                = leaf
  ...| (lt ^ g_1 |⌈ x_1 ⌉| rt ^ g_2) p_1 p_2 =
    (IH g_1 (sub_l g_2) lt |⌈ x_1 ⌉| IH g_2 (sub_r g_1) rt) p_1 p_2
```

We define apoS, the early-return variant of unfoldS, similarly. Note that no new proof obligations arise in the definition of apoS with respect to unfoldS.

The proof arguments of the datatype are irrelevant to the definition of unfoldS. This observation also means that the definition of unfold for the heap base functor H is syntactically identical. We consider it a question for future work whether the use of e.g. ornaments [McBride 2011], or possibly a folklore technique using higher-kinded types [Maguire 2018] could spare one this duplication.

### 5.3 Distributive Laws

Having defined the datatypes and requisite higher-order functions, we can define the distributive laws for the two-step bialgebraic sorting algorithms. They are structurally the same as the ones in [Hinze et al. 2012], but differ in that they additionally carry local proofs of element preservation and orderedness. We refer the reader to the formalization [Alexandru et al. 2024] for these detailed proofs.

Here, we focus instead on the fact that subcomponents of the bialgebraic sorting algorithms have uses in their own right. For that we first give an operational intuition of two-step bialgebraic sorting. We consider tree/quicksort in detail.

As both steps of bialgebraic semantics occurring in the tree-based algorithms yield two algorithms, two for growing and two for flattening a search tree, one can combine them to obtain four algorithms with distinct recursion behaviour. The way the search tree is built up determines whether we are dealing with tree sort or quicksort.

Quicksort builds the search tree coalgebraically. It outputs the pivot as the top node of the binary tree in each step. Its nested step, which partitions the list, is algebraic. It begins with the empty list and successively adds elements to the lists left or right of the pivot.

Treesort builds up the search tree algebraically. It successively inserts elements from the input into an initially empty tree. The nested step of treesort is coalgebraic, as insertion of an element into a tree successively outputs the nodes of the new tree until the proper insertion point is found.

#### 5.3.1 Intrinsically Correct Insertion.
The distributive law that these two variants arise from is sprout : ... → L ((S $_\times$) r) g → S ((L $_+$) r) g . We want to highlight that the nested step of treesort doubles as an intrinsically correct function for insertion into a binary tree.

treeSortStep : {g : FMSet A} → (L Stree) g → Stree g
treeSortStep {g} = apoS (sprout ∘ L_1 (⟨ id , outS ⟩ {g}))

Intrinsically correct insertion should take a pair of an element and a binary search tree, and return a search tree with this element inserted. If we index the pair by the element and the elements of the search tree, this can again be encoded as index preservation. Such a pair actually corresponds to the _::_ case of L. We introduce an auxiliary datatype for it:

data Cons (r : FMSet A → Type ℓ) : FMSet A → Type ℓ
  where _::_ : ∀ {g} → (x : A) → (r g) → Cons r (x :: g)

We obtain intrinsically correct insertion by adapting tree sort's nested step. Note that the :: constructor here is overloaded as a constructor of both Cons and L.

insert : {g : FMSet A} → (Cons Stree) g → Stree g
insert (a :: c2) = treeSortStep (a :: c2)

#### 5.3.2 Intrinsically Correct deleteMin.
Flattening a search tree into a list can again be done in the two ways. Coalgebraically, we extract the least element in each step. This extraction in turn is algebraic, as we start at the leaves and in each step return the candidate least element and the tree it was extracted from. Algebraically, we merge sorted sublists around their pivot. This operation is coalgebraic, as we incrementally output elements in order as we append the lists. The distributive law that these two variants arise from is called wither, and can be found in the Agda formalization [Alexandru et al. 2024].

A *deletemin* operation extracts the least element from a binary search tree. It is intrinsically correct if

- the element extracted is indeed smaller than all the elements
- the elements of the input tree are the elements of the output tree, plus the extracted element
- the output tree is still a search tree

The nested step of the coalgebraic variant of tree flattening is, without any adjustment, intrinsically correct.

deleteMin : {g : FMSet A} → Stree g → O Stree g
deleteMin {g} = (paraS ( O_1 ([ id , inS ] {g}) ∘ wither))

## 6 Ordered lists as a final coalgebra

In the previous sections we described how to formulate and prove intrinsic correctness of sorting algorithms in cubical Agda. In the current section we view our work through the lens of category theory, and in particular, we show how ordered lists can be obtained as a certain type of final coalgebra in a slice category over finite multisets. Detailed proofs can be found in Appendix A.

In order to see why using the slice category is relevant, we first note that the sorting algorithms as studied in Hinze

Intrinsically Correct Sorting in Cubical Agda

et al. [2012] are based on instantiating the framework of distributive laws and bialgebras [Klin 2011; Turi and Plotkin 1997] in the category Set of sets and functions. This instantiation mimics the development in Section 2, where it is written in terms of functional programs. We note that the Set-based 1-categorical semantics given in this section are compatible with our homotopy-type-theoretic formalization [Rijke and Spitters 2015].

Throughout this section, let $(A, \leq)$ be a totally ordered set of data elements whose lists we wish to sort. We are interested in the list functor, defined by $L \colon \mathsf{Set} \to \mathsf{Set}$, $L(X) = 1 + A \times X$, where $1 = \{\star\}$. As before, we alias this functor as $O = L$, and the *swap* operation then arises as a distributive law. This distributive law gives rise to a map

$$\mathsf{sort} \colon \mu L \to \nu L$$

from the initial $L$-algebra $\mu L$ to the final $O$-coalgebra $\nu O$, which is the actual operation of sorting, defined either via initiality or via finality. The initial algebra consists of the set $A^*$ of arbitrary lists, which makes sense, but the final coalgebra consists of all finite and infinite lists (e.g., [Jacobs 2016]). Moreover, these are not necessarily ordered.

We therefore show how to obtain the object of ordered lists as a final coalgebra in a different category. To this end, let $\mathcal{M}(A) = \{\varphi \colon A \to \mathbb{N} \mid \{x \mid \varphi(x) \neq 0\} \text{ is finite}\}$ be the set of finitely supported multisets over $A$. The *slice category* $\mathsf{Set}/\mathcal{M}(A)$ has as objects pairs $(X, f \colon X \to \mathcal{M}(A))$ where $X$ is a set and $f$ a function, and a morphism $h \colon (X, f) \to (Y, g)$ is a map $h \colon X \to Y$ such that $g \circ h = f$.

From a type theoretic perspective, we can understand objects of the slice category as dependent types. If we have a type $B$, then pairs of types $A$ and functions $f \colon A \to B$ are the same as a type families fib $: B \to$ Type. Families of sets fib $: B \to$ Set correspond to pairs of sets $A$ and functions $A \to B$. For this reason, we can view types dependent on multisets as functions into the type of multisets.

We define the functor $\hat{L} \colon \mathsf{Set}/\mathcal{M}(A) \to \mathsf{Set}/\mathcal{M}(A)$ by

$$\hat{L}(X, f) = (1 + A \times X, [\bar{0}, \lambda(a, x).\, \eta(a) \uplus f(x)])$$

where $\uplus$ is notation for multiset union, $\eta$ singleton inclusion, and $\bar{0}$ denotes the function mapping the single element of $1$ to the empty multiset. Let $\mathsf{elmts} \colon A^* \to \mathcal{M}(A)$ be the function that maps a list to the multiset of its elements (that is, forgetting about the order of elements). Note that the dependent type EIList in our Agda development corresponds to the slice map $\mathsf{elmts} \colon A^* \to \mathcal{M}(A)$. The initial algebra of $\hat{L}$ is essentially that of $L$.

**Lemma 6.1.** *The object* $(A^*, \mathsf{elmts})$ *is the carrier of the initial algebra for* $\hat{L}$.

Perhaps more surprisingly, the fact that we work in the slice category over $\mathcal{M}(A)$ ensures an *initial algebra/final coalgebra coincidence*. Where the final $L$-coalgebra consists of lists and (infinite) streams, the final $\hat{L}$-coalgebra contains only finite lists.

**Theorem 6.2.** *The object* $(A^*, \mathsf{elmts})$ *is the carrier of the final coalgebra for* $\hat{L}$.

To see why the final coalgebra consists only of finite lists, let $c \colon (X, f) \to \hat{L}(X, f)$ be a coalgebra. The key observation is that, if $c(x) = (a, x')$, then $f(x')(a) = f(x)(a) - 1$, i.e., the size of the multiset of the next state is strictly smaller than that of the current state. As a consequence, iterating the coalgebra on a given state $x \in X$ can only happen finitely many times before reaching the element $\star \in 1$. Notice that this argument relies on the multisets having finite support.

Using the slice maps, we can also define an ordered variant of $L$. Let $\hat{O} \colon \mathsf{Set}/\mathcal{M}(A) \to \mathsf{Set}/\mathcal{M}(A)$ be given by:

$\hat{O}(X, f) = (1 + A \times_\leq X, [\bar{0}, \lambda(a, x).\, \eta(a) \uplus f(x)])$, where $A \times_\leq X = \{(a, x) \in A \times X \mid a \leq \min\{b \mid f(x)(b) \neq 0\}\}$.

Let $A^*_\leq$ be the set of ordered lists. The type OEIList in our earlier Agda development corresponds to the slice map $\mathsf{elmts} \colon A^*_\leq \to \mathcal{M}(A)$. We arrive at the main observation of this section:

**Theorem 6.3.** *The object* $(A^*_\leq, \mathsf{elmts})$ *is the carrier of a final coalgebra for* $\hat{O}$.

This result encapsulates two properties that are essential for our purposes. First, the final coalgebra consists only of *finite* lists (as in Theorem 6.2), and second, they are ordered. The orderedness is enforced by the use of the restriction $A \times_\leq X$ in the definition of $\hat{O}$.

In conclusion, maps from the initial $\hat{L}$-algebra to the final $\hat{O}$-coalgebra are of the following form:

$$\begin{array}{ccc} A^* & \xrightarrow{\mathsf{sort}} & A^*_\leq \\ & \searrow^{\mathsf{elmts}} \quad \swarrow_{\mathsf{elmts}} & \\ & \mathcal{M}(A) & \end{array}$$

This is precisely the type of intrinsically correct sorting algorithms: they turn arbitrary lists into ordered lists, with the same elements, thanks to the commutativity of the above diagram.

*Remark.* The essence of the above result lies in an initial-algebra/final-coalgebra coincidence, enforced by the slicing over finite multisets. The recent [Kori et al. 2021] provides a systematic study of such coincidences in a fibrational setting. The coincidence there occurs in a specific fibre, as opposed to the total category as above. We leave a logical/fibrational treatment of the above development for future work. Note that this perspective is not immediately clear, as the functor $\hat{O}$ is not a lifting of $O$ (it is not possible to express orderedness in the base category).

## 7 Related Work

The insight that the sorting algorithms insertion/selection sort, tree/quicksort and heapsort can be expressed as two algorithms for the price of one shared business logic in the form of a distributive law is due to Hinze et al. [2012]. However, they did not prove the correctness or termination of these algorithms. For the analysis of intrinsic correctness in the current paper, since bialgebraic semantics necessarily makes use of final coalgebra semantics, it was necessary to prove that all coalgebras constructed are terminating. We proved this more generally, by showing it is the case for *all* coalgebras for the element-indexed ordered list / tree base functors we defined. This went hand-in-hand with our intrinsic verification of the correctness of these algorithms.

With regards to the intrinsic encoding of orderedness invariants, McBride [2014, Section 4] presents an approach of defining datatypes with orderedness invariants by having them be indexed by bounds, and having constructors respect those bounds. These bound requirements are then pushed inward, instead of being "measurements" that are propagated outward. We stuck with the more traditional "measurement" approach, as the multiset index was already required for the element preservation property and it served a double use in encoding the orderedness invariant.

Kupke et al. [2023] use "fresh lists" to define ordered lists where elements to be prepended are required to be smaller than all the elements in the list. This is not done at the level of base functors, however, but only in the definition of the inductive datatype, and as such doesn't use indexing. In fact, they show that ordered lists are a possible *implementation* of finite multisets.

On the topic of element-preservation, Danielsson [2012] defines multiset equivalence in Agda, and as an example application extrinsically verifies that a tree-sort preserves elements.

Appel [2017] provides a verification of insertion- and selection-sort, using multisets modeled as both maps of type $\mathbb{N} \to A$ and lists up to bag equivalence. They also implicitly encounter the issue of selection sort being coalgebraic in nature and thus not structurally recursive on the list; they address this by providing the length of the list as a fuel parameter. This differs from our approach in that the length needs to be computed and passed as a parameter explicitly. The main difference is that the approach in *op. cit.* is extrinsic, as opposed to the intrinsic verification approach in the current paper.

The approach of ensuring element preservation by means of indexing by a multiset has been experimented with before by Atkey [2013]. The approach there was to develop an embedded domain-specific language for a subset of the linear lambda calculus where a context keeps track of a list of terms, which substructural rules allow to be treated as a multiset. However, this approach means that algorithms are expressed as terms in a deeply-embedded DSL, whose *interpretation* yields a sorting algorithm in the host language, which is a level of indirection not present in our case.

## 8 Discussion

Throughout the development, we used inductive types indexed by higher inductive types. While this gives us a nice and convenient way to define EIList and OEIList, it does come with some disadvantages. Concretely, the normal forms of types indexed by HITs are only given by the constructors. As a consequence, the algorithms that we described in our paper do not necessarily compute to a list. The term insertSort (2 ∷ 1 ∷ []), for example, does not compute to 1 ≤∷ 2 ≤∷ [], but to a transport.

To understand why, let us consider a simple example of a type indexed by a HIT.

data I : Type where
  o   : I
  i    : I
  seq : o ≡ i

data T : I → Type where
  x : T o

If we try to normalize the term subst T seq x, then cubical Agda returns the following term.

$$\text{transpX-T } (\lambda\ n \to \ldots)\ i0\ x$$

This result is fair. Since we did not specify any constructor of T i, there is no constructor to which it even could evaluate. This behavior corresponds to what has been described by Cavallo and Harper [2019]. They described the canonical normal forms of types indexed by higher inductive types as either a constructor, a coercion, or a composition.

To obtain a term of a normal form that isn't a transport, we have to convert the ordered, element-indexed list back into an ordinary list. To this end we define the following conversion.

OEIListToList : {g : FMSet A} → OEIList g → List A
OEIListToList [] = []
OEIListToList ((x ≤∷ xs) _) = x ∷ OEIListToList xs

Then, OEIListToList (insertSort (2 ∷ 1 ∷ [])) normalizes to 1 ∷ 2 ∷ []. However, this could be considered a workaround.

If we want transports for element-indexed lists to reduce, there are several approaches we could take. One of these would be to define element-indexed lists in a different way. Instead of defining this type as an indexed inductive type, we could also define it using recursion on finite multisets. This avoid the usage of indexed indexed types, but it does come at another cost. First, we are required to use the free symmetric monoidal groupoid [Piceghello 2020, 2021] instead. This is because the recursion principle of finite multisets only allows us to eliminate into a type that is a set. Since the type of sets is not a set itself, we cannot map into it using the recursion principle. The free symmetric monoidal groupoid, on the other hand, is defined to be a *groupoid* rather than a set, and for that reason, we can use recursion to map into the type of sets. The second cost is that this alternative definition is more convoluted and less readable.

Another possible solution would require further study of types indexed by higher inductive types. If we define a type indexed by a HIT, then their canonical normal forms are not fully specified, because transports also are canonical normal forms. One could view this as some kind of underspecification of the indexed type, since the transport functions are not fully specified, in the sense that they do not always reduce further. A version of indexed types in which one could also specify what the transport functions reduce to would be able to overcome this hindrance. We leave the study of such types as future work.

## 9 Conclusion and Future Work

We have proposed a way to define intrinsic correctness of sorting algorithms in cubical Agda. To this end we made use of types indexed by finite multisets, encoded as a quotient inductive type. With this foundation we have shown how, in the bialgebraic approach to sorting algorithms, verifying the basic underlying distributive laws that contain the essential business logic suffices to ensure correctness of the ensuing sorting algorithms. In particular, we have verified insertion/selection sort as a pair of algorithms generated by a simple distributive law, and treesort/quicksort and heapsort as more elaborate examples.

There are several avenues for future work.

***Recursive and well-founded coalgebras.*** From a coalgebraic perspective, the main new idea is that indexing over a multiset yields finite lists as a final coalgebra. The final coalgebra thereby becomes *well-founded*, suggesting a link to the theory of (co)recursive and well-founded coalgebras [Taylor 1999]. It would be interesting to further explore the possible role of these notions in bialgebraic semantics.

***Beyond sorting.*** The theory of bialgebras and distributive laws was originally proposed by Turi and Plotkin in the context of operational semantics, where distributive laws generalise the celebrated GSOS rule format [Klin 2011; Turi and Plotkin 1997]. They have been used as well in (co)algebraic presentations of automata constructions [Silva et al. 2013]. It remains open to investigate similar indexing as proposed in this paper in such examples, possibly yielding proof methods for finiteness of such calculi and constructions at a high level of generality.

## Acknowledgments

The authors would like to thank Ruben Turkenburg, Bálint Kocsis, and Bob Atkey for discussions. This research is supported by the EU Marie-Skłodowska-Curie action ReGraDe-CS, grant № 101106046 (https://doi.org/10.3030/101106046); the NWO project "The Power of Equality" OCENW.M20.380 (https://www.nwo.nl/projecten/ocenwm20380); and the NWO project OCENW.M20.053 (https://www.nwo.nl/projecten/ocenwm20053). The latter two projects are both financed by the Dutch Research Council (NWO).

## References

The Agda Team. 2024. *Function Definitions : Dot Patterns — Agda 2.7.0.1 documentation*. https://agda.readthedocs.io/en/v2.7.0.1/language/function-definitions.html#dot-patterns

Cass Alexandru, Vikraman Choudhury, Jurriaan Rot, and Niels van der Weide. 2024. *Intrinsically Correct Sorting in Cubical Agda*. https://doi.org/10.5281/zenodo.14279034

Andrew Appel. 2017. *Verified Functional Algorithms* (1.5.5 ed.). Software Foundations, Vol. 3. https://softwarefoundations.cis.upenn.edu/vfa-1.5.5/

Bob Atkey. 2013. Typed DSLs for Sorting. https://github.com/bobatkey/sorting-types

Gilles Barthe, Venanzio Capretta, and Olivier Pons. 2003. Setoids in Type Theory. *Journal of Functional Programming* 13, 2 (March 2003), 261–293. https://doi.org/10.1017/S0956796802004501

Richard S. Bird and Oege de Moor. 1997. *Algebra of Programming*. Prentice Hall.

Richard J. Boulton, Andrew D. Gordon, Michael J. C. Gordon, John Harrison, John Herbert, and John Van Tassel. 1992-06-22/1992-06-24. Experience with Embedding Hardware Description Languages in HOL. In *Proceedings of the IFIP TC10/WG 10.2 International Conference on Theorem Provers in Circuit Design: Theory, Practice and Experience (IFIP Transactions, Vol. A-10)*, Victoria Stavridou, Thomas F. Melham, and Raymond T. Boute (Eds.). North-Holland, Nijmegen, The Netherlands, 129–156. https://doi.org/10.5555/645902.672777

Evan Cavallo and Robert Harper. 2019. Higher Inductive Types in Cubical Computational Type Theory. *Proc. ACM Program. Lang.* 3, POPL (Jan. 2019), 1:1–1:27. https://doi.org/10.1145/3290314

Vikraman Choudhury and Marcelo Fiore. 2023. Free Commutative Monoids in Homotopy Type Theory. *Electronic Notes in Theoretical Informatics and Computer Science* Volume 1 - Proceedings of MFPS XXXVIII (Feb. 2023). https://doi.org/10.46298/entics.10492

Cyril Cohen, Thierry Coquand, Simon Huber, and Anders Mörtberg. 2017. Cubical Type Theory: A Constructive Interpretation of the Univalence Axiom. *FLAP* 4, 10 (2017), 3127–3170.

Nils Anders Danielsson. 2012. Bag Equivalence via a Proof-Relevant Membership Relation. In *Interactive Theorem Proving (Lecture Notes in Computer Science)*, Lennart Beringer and Amy Felty (Eds.). Springer, Berlin, Heidelberg, 149–165. https://doi.org/10.1007/978-3-642-32347-8_11

Stijn de Gouw, Frank S. de Boer, Richard Bubel, Reiner Hähnle, Jurriaan Rot, and Dominic Steinhöfel. 2019. Verifying OpenJDK's Sort Method for Generic Collections. *J. Autom. Reason.* 62, 1 (2019), 93–126.

Peter Dybjer. 1994. Inductive Families. *Formal Aspects of Computing* 6, 4 (July 1994), 440–465. https://doi.org/10.1007/BF01211308

Dan Frumin, Herman Geuvers, Léon Gondelman, and Niels van der Weide. 2018. Finite Sets in Homotopy Type Theory. In *Proceedings of the 7th ACM SIGPLAN International Conference on Certified Programs and Proofs (CPP 2018)*. Association for Computing Machinery, New York, NY, USA, 201–214. https://doi.org/10.1145/3167085

Jeremy Gibbons and Nicolas Wu. 2014. Folding Domain-Specific Languages: Deep and Shallow Embeddings (Functional Pearl). In *Proceedings of the 19th ACM SIGPLAN International Conference on Functional Programming* (ICFP '14). Association for Computing Machinery, New York, NY, USA, 339–347. https://doi.org/10.1145/2628136.2628138

Claudio Hermida and Bart Jacobs. 1995. An Algebraic View of Structural Induction. In *Computer Science Logic*, Leszek Pacholski and Jerzy Tiuryn (Eds.). Springer, Berlin, Heidelberg, 412–426. https://doi.org/10.1007/BFb0022272

Ralf Hinze, Daniel W. H. James, Thomas Harper, Nicolas Wu, and José Pedro Magalhães. 2012. Sorting with Bialgebras and Distributive Laws. In *Proceedings of the 8th ACM SIGPLAN Workshop on Generic Programming, WGP@ICFP 2012, Copenhagen, Denmark, September 9-15, 2012*, Andres Löh and Ronald Garcia (Eds.). ACM, 69–80. https://doi.org/10.1145/2364394.2364405

Bart Jacobs. 2016. *Introduction to Coalgebra: Towards Mathematics of States and Observation*. Number 59 in Cambridge Tracts in Theoretical Computer Science. Cambridge university press, Cambridge.

JetBrains Research. 2021. *Universes – Arend Theorem Prover documentation*. https://arend-lang.github.io/documentation/language-reference/expressions/universes.html

Bartek Klin. 2011. Bialgebras for Structural Operational Semantics: An Introduction. *Theor. Comput. Sci.* 412, 38 (2011), 5043–5069.

Mayuko Kori, Ichiro Hasuo, and Shin-ya Katsumata. 2021. Fibrational Initial Algebra-Final Coalgebra Coincidence over Initial Algebras: Turning Verification Witnesses Upside Down. In *CONCUR (LIPIcs, Vol. 203)*. Schloss Dagstuhl - Leibniz-Zentrum für Informatik, 21:1–21:22.

Clemens Kupke, Fredrik Nordvall Forsberg, and Sean Watters. 2023. A Fresh Look at Commutativity: The 21st Asian Symposium on Programming Languages and Systems. 1–20.

Marina Lenisa, John Power, and Hiroshi Watanabe. 2000. Distributivity for Endofunctors, Pointed and Co-Pointed Endofunctors, Monads and Comonads. In *Coalgebraic Methods in Computer Science, CMCS 2000, Berlin, Germany, March 25-26, 2000 (Electronic Notes in Theoretical Computer Science, Vol. 33)*, Horst Reichel (Ed.). Elsevier, 230–260. https://doi.org/10.1016/S1571-0661(05)80350-0

Sandy Maguire. 2018. Higher-Kinded Data. https://reasonablypolymorphic.com/blog/higher-kinded-data/

Maria Emilia Maietti. 1998. About Effective Quotients in Constructive Type Theory. In *Types for Proofs and Programs, International Workshop TYPES '98, Kloster Irsee, Germany, March 27-31, 1998, Selected Papers (Lecture Notes in Computer Science, Vol. 1657)*, Thorsten Altenkirch, Wolfgang Naraschewski, and Bernhard Reus (Eds.). Springer, 164–178. https://doi.org/10.1007/3-540-48167-2_12

P. Martin-Löf. 1984. Constructive Mathematics and Computer Programming. In *Philosophical Transactions of the Royal Society of London. Series A. Mathematical and Physical Sciences*. Vol. 312. 501–518. https://doi.org/10.1098/rsta.1984.0073

Conor McBride. 2011. Ornamental Algebras, Algebraic Ornaments. (2011). https://personal.cis.strath.ac.uk/conor.mcbride/pub/OAAO/Ornament.pdf

Conor McBride. 2014. How to Keep Your Neighbours in Order. In *Proceedings of the 19th ACM SIGPLAN International Conference on Functional Programming*. ACM, Gothenburg Sweden, 297–309. https://doi.org/10.1145/2628136.2628163

Conor McBride and James McKinna. 2004. The view from the left. *Journal of Functional Programming* 14, 1 (Jan. 2004), 69–111. https://doi.org/10.1017/S0956796803004829

Lambert Meertens. 1992. Paramorphisms. *Formal Aspects of Computing* 4, 5 (Sept. 1992), 413–424. https://doi.org/10.1007/BF01211391

Tobias Nipkow, Jasmin Blanchette, Manuel Eberl, Alejandro Gómez-Londoño, Peter Lammich, Christian Sternagel, Simon Wimmer, and Bohua Zhan. 2021. *Functional Algorithms, Verified!*

Stefano Piceghello. 2020. Coherence for Monoidal Groupoids in HoTT. In *DROPS-IDN/v2/Document/10.4230/LIPIcs.TYPES.2019.8*. Schloss Dagstuhl – Leibniz-Zentrum für Informatik. https://doi.org/10.4230/LIPIcs.TYPES.2019.8

Stefano Piceghello. 2021. *Coherence for Monoidal and Symmetric Monoidal Groupoids in Homotopy Type Theory*. Doctoral Thesis. The University of Bergen. https://bora.uib.no/bora-xmlui/handle/11250/2830640

Andrew M. Pitts. 2020. *The combination of Cubical Agda with inductive families is logically inconsistent · Issue #4606 · agda/agda*. https://github.com/agda/agda/issues/4606

Egbert Rijke and Bas Spitters. 2015. Sets in homotopy type theory. *Mathematical Structures in Computer Science* 25, 5 (June 2015), 1172–1202. https://doi.org/10.1017/S0960129514000553

Alexandra Silva, Filippo Bonchi, Marcello M. Bonsangue, and Jan J. M. M. Rutten. 2013. Generalizing Determinization from Automata to Coalgebras. *Log. Methods Comput. Sci.* 9, 1 (2013).

M. B. Smyth and G. D. Plotkin. 1982. The Category-Theoretic Solution of Recursive Domain Equations. *SIAM J. Comput.* 11, 4 (Nov. 1982), 761–783. https://doi.org/10.1137/0211062

Paul Taylor. 1999. *Practical Foundations of Mathematics*. Number 59 in Cambridge Studies in Advanced Mathematics. Cambridge university press, Cambridge.

The agda/cubical development team. 2018/. The Agda/Cubical Library. https://github.com/agda/cubical/

Daniele Turi and Gordon D. Plotkin. 1997. Towards a Mathematical Operational Semantics. In *Proceedings, 12th Annual IEEE Symposium on Logic in Computer Science, Warsaw, Poland, June 29 - July 2, 1997*. IEEE Computer Society, 280–291. https://doi.org/10.1109/LICS.1997.614955

D. A. Turner. 1995. Elementary Strong Functional Programming. In *Funtional Programming Languages in Education*, Pieter H. Hartel and Rinus Plasmeijer (Eds.). Springer, Berlin, Heidelberg, 1–13. https://doi.org/10.1007/3-540-60675-0_35

The Univalent Foundations Program. 2013. *Homotopy Type Theory: Univalent Foundations of Mathematics*. Institute for Advanced Study. https://homotopytypetheory.org/book

Varmo Vene and Tarmo Uustalu. 1998. Functional Programming with Apomorphisms (Corecursion). *Proceedings of the Estonian Academy of Sciences. Physics. Mathematics* 47, 3 (1998), 147. https://doi.org/10.3176/phys.math.1998.3.01

Andrea Vezzosi, Anders Mörtberg, and Andreas Abel. 2019. Cubical Agda: A Dependently Typed Programming Language with Univalence and Higher Inductive Types. *Proc. ACM Program. Lang.* 3, ICFP (July 2019), 87:1–87:29. https://doi.org/10.1145/3341691

## A  Proofs for Section 6

### A.1  Initial Algebra

Consider the following functions:

$$\begin{aligned}
\mathsf{cons} &: A \times A^* \to A^* \\
\mathsf{cons} &:= \lambda(a, \langle a_0, ..., a_{n-1}\rangle). \langle a, a_0, ..., a_{n-1}\rangle \\
\overline{\langle\rangle} &: 1 \to A^* \\
\overline{\langle\rangle} &:= \lambda \star . \langle\rangle \\
\mathsf{in} &: LA^* \to A^* \\
\mathsf{in} &:= [\overline{\langle\rangle}, \mathsf{cons}]
\end{aligned}$$

**Lemma A.1.** $(A^*, \mathsf{in})$ *is the initial L-algebra.*

*Proof.* Let $(X, [n, c])$ be some $L$-algebra. Then we must show there is a unique algebra morphism $f$ from $[\overline{\langle\rangle}, \mathsf{cons}]$ to $[n, c]$,

s.t. the following diagram commutes:

$$\begin{array}{ccc} LA^* & \xrightarrow{Lf} & LX \\ {\scriptstyle [\langle\rangle,\mathsf{cons}]}\downarrow & & \downarrow{\scriptstyle [n,c]} \\ A^* & \xdashrightarrow[\exists f!]{} & X \end{array}$$

This is requirement is expressed as the following equalities:

$$f(\langle\rangle) = n(\star)$$
$$f(\mathsf{cons}(a,r)) = c(a, f(r))$$

We define $f$ as the unique solution to this system of equations. □

**Notation A.1.** For a functor $F$ with an initial algebra $\mathsf{in}_F$, we write $(\!(a)\!)$ for the unique morphism from $\mathsf{in}_F$ to some target $F$-algebra $a$.

We use the initiality just proven to define a function which maps a list to the multiset of its elements:

**Definition A.1.** Let $\mathsf{elt} := [\bar{0}, \lambda(a,r).\,\eta(a) \uplus r]$ be the algebra for multiset insertion. Then elt uniquely extends to an $L$-algebra morphism $\mathsf{elmts} := (\!(\mathsf{elt})\!): A^* \to \mathcal{M}(A)$ which maps a list to the multiset of its elements.

$$\begin{array}{ccc} LA^* & \xrightarrow{L(\mathsf{elt})} & L(\mathcal{M}(A)) \\ {\scriptstyle [\langle\rangle,\mathsf{cons}]}\downarrow & & \downarrow{\scriptstyle \mathsf{elt}=[\bar{0},\uplus\circ(\eta\times\mathsf{id})]} \\ A^* & \xdashrightarrow{(\!(\mathsf{elt})\!)} & \mathcal{M}(A) \end{array}$$

*Remark.* We write $\lambda(a,r).\,\eta(a) \uplus r$ pointfree as $\uplus \circ (\eta \times \mathsf{id})$.

**Lemma** (6.1). $((A^*, \mathsf{elmts}), \mathsf{in})$ *is the initial* $\hat{L}$*-algebra.*

*Proof.* We note that there is an isomorphism of categories $\mathsf{Alg}(\hat{L}) \simeq \mathsf{Alg}(L)/\mathsf{elt}$. Namely, given an algebra $\hat{L}(X,g) \xrightarrow{a} (X,g)$, the slice map of its domain, $[\bar{0}, \uplus \circ (\eta \times g)]$ is equivalently $\mathsf{elt} \circ Lg$, and $g$ is an $L$-algebra-morphism from $a$ to elt:

$$\begin{array}{ccc} LX & \xrightarrow{a} & X \\ {\scriptstyle Lg}\downarrow & {\scriptstyle [\bar{0},\uplus\circ(\eta\times g)]}\searrow & \downarrow{\scriptstyle g} \\ L(\mathcal{M}(A)) & \xrightarrow[\mathsf{elt}=[\bar{0},\uplus\circ(\eta\times\mathsf{id})]]{} & \mathcal{M}(A) \end{array}$$

This isomorphism just exchanges slice map- and morphism components. We thus proceed to prove that $((A^*, \mathsf{in}), (\!(\mathsf{elt})\!))$ is initial in $\mathsf{Alg}(L)/\mathsf{elt}$.

Let $((X,a),g)$ be an object in $\mathsf{Alg}(L)/\mathsf{elt}$. We must construct a unique $L$-algebra morphism $f$ from $((A^*, \mathsf{in}), (\!(\mathsf{elt})\!))$ to $((X,a),g)$, s.t. the following diagram commutes:

$$\begin{array}{c} L(\mathcal{M}(A)) \\ {\scriptstyle L(\mathsf{elt})}\nearrow \quad \downarrow \quad \nwarrow{\scriptstyle Lg} \\ LA^* \xdashrightarrow{Lf} LX \\ {\scriptstyle \mathsf{in}}\downarrow \quad {\scriptstyle \mathsf{elt}}\downarrow \quad \downarrow{\scriptstyle a} \\ \quad \mathcal{M}(A) \\ {\scriptstyle (\!(\mathsf{elt})\!)}\nearrow \quad \nwarrow{\scriptstyle g} \\ A^* \xrightarrow{f} X \end{array}$$

We define $f := (\!(a)\!)$ by initiality of $(A^*, \mathsf{in})$. It remains to prove that $(\!(a)\!)$ is a slice morphism, i.e. $g \circ (\!(a)\!) = (\!(\mathsf{elt})\!)$. This follows from the uniqueness of $(\!(\mathsf{elt})\!): (A^*, \mathsf{in}) \to (\mathcal{M}(A), \mathsf{elt})$, since $g \circ (\!(a)\!)$ is an $L$-algebra morphism of the same type. □

### A.2 Final Coalgebra

**Theorem** (6.2). $((A^*, \mathsf{elmts}), \mathsf{in}^{-1})$ *is the final* $\hat{L}$*-coalgebra.*

*Proof.* We define the following function length that gives us the number of elements of a finite multiset:

$$\mathsf{length}\colon \mathcal{M}(A) \to \mathbb{N}$$
$$\mathsf{length}(\varphi) := \sum_{a \in A} \varphi(a)$$

This sum is well defined since $\mathcal{M}(A)$ has finite support.

We define the following relation $x \prec y := \mathsf{length}(x) < \mathsf{length}(y)$. This relation is well founded as the inverse image under length of the well founded relation $<$.

Consider an $L$-coalgebra $c\colon (X,g) \to \hat{L}(X,g)$.

$$\begin{array}{ccc} X & \xrightarrow{c} & L(X) \\ {\scriptstyle g}\searrow & & \swarrow{\scriptstyle [\uplus\circ(\eta\times g),\bar{0}]} \\ & \mathcal{M}(A) & \end{array} \quad (1)$$

We want to define a unique coalgebra morphism to $\mathsf{in}^{-1}$. Note that this is equivalent to defining a coalgebra-to-algebra morphism to in, since by Lambek's lemma, in is an isomorphism. So we must construct a unique morphism $f$ s.t. the following diagram commutes.

$$\begin{array}{ccc} LX & \xdashrightarrow{Lf} & LA^* \\ {\scriptstyle \hat{L}g}\searrow \; \nearrow{\scriptstyle \hat{L}\mathsf{elmts}} & & \\ {\scriptstyle c}\uparrow & \mathcal{M}(A) & \uparrow{\scriptstyle [\langle\rangle,\mathsf{cons}]} \\ & {\scriptstyle g}\nearrow \; \nwarrow{\scriptstyle \mathsf{elmts}} & \\ X & \xdashrightarrow{f} & A^* \end{array} \quad (2)$$

We do this by switching from the sliced to the indexed perspective and defining $f$ as a family of maps in Set between the fibres of $g$ and elmts, using well founded recursion.

$$(f_x \colon g^{-1}(x) \to \mathsf{elmts}^{-1}(x))_{x \in \mathcal{M}(A)}$$

Concretely, well founded recursion gives us a unique such family of maps $(f_x)_{x \in \mathcal{M}(A)}$, provided we can construct $f_x$ given $(f_y)_{y \in \{z \mid z \prec x\}}$. We define $f_x$ as follows.

$$f_x \colon g^{-1}(x) \to \mathsf{elmts}^{-1}(x)$$

$$f_x(s) := \begin{cases} \langle\rangle & c(s) = \star \\ \mathsf{cons}(a, f_{g(s')}(s')) & c(s) = (a, s') \end{cases}$$

It remains to show $g(s') \prec x$. From $s \in g^{-1}(x) \Rightarrow x = g(s)$ and $g(s) \stackrel{(1)}{=} \eta(a) \uplus g(s')$ we have $\mathsf{length}(g(s')) < \mathsf{length}(\eta(a) \uplus g(s')) = \mathsf{length}(g(s)) = \mathsf{length}(x)$.

It is unique as the unique function making diagram (2) commute (proof by diagram chase). □

**Definition A.2** (Ordered Lists). Ordered lists are lists whose consecutive elements are related by $\leq$.

$$A^*_\leq := \{\langle a_0, \ldots, a_{n-1}\rangle \in A^* \mid \bigwedge_{i=0}^{n-2} a_i \leq a_{i+1}\}$$

**Lemma A.2.** *Let $x = \langle a_0, \ldots, a_{n-1}\rangle$ in:*

$$\min\{b \mid \mathsf{elmts}(x)(b) \neq 0\} = \min\{a_0, \ldots, a_{n-1}\}$$

*Proof.* By induction on $x$ □

**Theorem** (6.3). $((A^*_\leq, \mathsf{elmts}), \mathsf{in}^{-1})$ *is the final $\hat{O}$-coalgebra.*

*Proof.* The wellfoundedness argument is the same as the one above. We additionally need to check that $f_x \colon g^{-1}(x) \to \mathsf{elmts}^{-1}(x)$ respects the refinement of $A^*$ to $A^*_\leq$. Consider some $\hat{O}$-coalgebra $c \colon (X, g) \to (1 + A \times_\leq X, [\bar{0}, \uplus \circ (\eta \times g)])$. We examine the clauses of $f$:

$$f_x \colon g^{-1}(x) \to \mathsf{elmts}^{-1}(x)$$

$$f_x(s) := \begin{cases} \langle\rangle & c(s) = \star \\ \mathsf{cons}(a, f_{g(s')}(s')) & c(s) = (a, s') \end{cases}$$

Let $A^*_\leq \ni \langle a_0, \ldots, a_{n-1}\rangle = f_{g(s')}(s')$. In the second clause of $f$, we know $a \leq \min\{b \mid g(s')(b) \neq 0\}$. But since $\mathsf{elmts}(f_{g(s')}(s')) = g(s')$, and thus (and by Lemma A.2), $\min\{a_0, \ldots, a_{n-1}\} = \min\{b \mid g(s')(b) \neq 0\}$, we have $a \leq \min\{a_0, \ldots, a_{n-1}\}$. Therefore, the application of cons is well defined. □